\documentclass[a4paper]{article}
\usepackage{INTERSPEECH2022}
\usepackage{mathtools}
\DeclarePairedDelimiter{\ceil}{\lceil}{\rceil}
\usepackage{multirow}
\usepackage{multicol}
\usepackage{array}
\usepackage{url}
\newcolumntype{A}{>{\centering}p{5mm}}
\title{Efficient Non-Autoregressive GAN Voice Conversion using VQWav2vec Features and Dynamic Convolution}
\name{Mingjie Chen$^1$, Yanghao Zhou$^2$, Heyan Huang$^3$, Thomas Hain$^1$}
\address{
  $^1$Department of Computer Science, University of Sheffield\\
  $^2$Southeast Academy of Information Technology, $^3$School of Computer Science,
  Beijing Institute of Technology}
\email{mchen33@sheffield.ac.uk, zhouyh77@bit.edu.cn, hhy63@bit.edu.cn, t.hain@sheffield.ac.uk}

\begin{document}

\maketitle

\begin{abstract}
%This paper proposes a novel generative adversarial network (GAN) voice conversion (VC) model with high efficiency. Most of the state-of-the-art (SOTA) VC models depend on large model size and autoregressive (AR) models. These limit the efficiency of VC systems in practical situations. This paper proposes Dynamic GAN VC (DYGAN-VC), aiming at improving efficiency by establishing a lightweight non-AR GAN model. DYGAN-VC uses VQWav2vec, which is a lightweight replacement of commonly used large AR speech recognition models. Besides, DYGAN-VC uses dynamic convolution which enhances speech content modeling and also keeps the model lightweight. DYGAN-VC has nearly half of parameters of the SOTA models. Furthermore, comparing to traditional Cascade ASR-TTS, DYGAN-VC has much faster decoding speed. Objective and subjective evaluation results show that DYGAN-VC reaches comparable performance of SOTA with higher efficiency %Experiment results show DYGAN-VC achieves comparable MOS scores 3.81 and 3.87 for naturalness and speaker similarity, respectively.
It was shown recently that a combination of ASR and TTS models  yield highly 
competitive performance on standard voice conversion tasks such as the Voice
Conversion Challenge 2020 (VCC2020). To obtain good performance
both models require pretraining on large amounts of data, thereby obtaining
large models that are potentially inefficient in use. In this work we present a model that
is significantly smaller and thereby faster in processing while obtaining equivalent performance. 
To achieve this the proposed model, Dynamic-GAN-VC (DYGAN-VC), uses a non-autoregressive structure
and makes use of vector quantised embeddings obtained from a VQWav2vec model. Furthermore 
dynamic convolution is introduced to improve speech content modeling while requiring a small
number of parameters. Objective and subjective evaluation was performed using the VCC2020 task, 
yielding MOS scores of up to 3.86, and character error rates as low as 4.3\%. This was achieved with approximately half the number of model parameters, and up to 8 times faster decoding speed. 

% The state-of-the-art (SOTA) voice conversion (VC) models usually are composed of an automatic speech recognition (ASR) model and an text-to-speech (TTS) model. Recently, autoregressive (AR) ASR and TTS models have been shown having large model size and slow decoding issues, which become limitations of cascade ASR-TTS VC systems in practical situations where memory and computation resources are limited.

% In order to improve parameter and decoding efficiency, this paper proposes Dynamic-GAN-VC (DYGAN-VC), which is a non-AR model with a smaller model size and faster decoding speed. 

% More specifically, instead of using an ASR model, this paper uses VQWav2vec, which has a smaller size.
% Furthermore, in order to improve parameter efficiency of convolution layers, dynamic convolution is introduced to the model. 

% DYGAN-VC has nearly half of the size of the cascade ASR-TTS model and also nearly 9 times faster decoding speed. 

% Objective and subjective evaluation results show that DYGAN-VC reaches comparable performance of the cascade ASR-TTS model with higher parameter and decoding efficiency.

\end{abstract}

\section{Introduction}
Recently, the state-of-the-art (SOTA) voice conversion (VC) models \cite{Huang2020,li2021starganv2,zhang20_vccbc,liu20_vccbc,ma20_vccbc,huang2019voice} have achieved good performance, the generated samples have reached near to human level of voice quality. 
In the recent Voice Conversion Challenge 2020 \cite{Yi2020} (VCC2020) Cascade ASR-TTS \cite{Huang2020} obtains competitive performance. It is composed of an automatic speech recognition (ASR) model and a text-to-speech (TTS) model. Both ASR and TTS models are large pretrained autoregressive (AR) Transformer \cite{vaswani2017attention} models. Recently, there have been several works showing that Transformer models have parameter and decoding efficiency issues in both ASR \cite{tomanek2021residual,song2021non} and TTS \cite{ihm2020reformer,luo2021lightspeech} areas.
Hence it would be inefficient to deploy a cascade ASR-TTS model in practical situations where memory and computation resources are limited. %Cascade ASR-TTS models For example, large AR VC models \cite{Huang2020,huang2019voice,zhang20_vccbc} are difficult to deploy in real time voice conversion because of their large size and the computational costs.

\begin{table*}[h]
    \centering
    \begin{tabular}{c|c|c|c|c}
     Model  & Text & Direction & \# Param & AR  \\
     \hline
     Cascade ASR-TTS \cite{Huang2020} &  Yes & Many-to-One & 78M(ASR)+25M(TTS) + 2M (Emb)&Yes \\
     %StarGANv2-VC \cite{li2021starganv2}  & Yes & Many-to-Many& 70M(GAN) + 12M(ASR) & No \\
     %CycleGAN-VC3 \cite{kaneko2020CycleGAN-VC3} & No & No & One2one & 27M(GAN)  & No \\
     %CycleVQVAE & No & No & One2one & 2M  & No \\
     DYGAN-VC (ours)&No& Many-to-Many & 7M(GAN) + 35M (VQWav2vec) + 1M(Emb)& No\\
     \hline
    \end{tabular}
    \caption{A comparison of VC models. The columns means: (1) use text transcriptions, (2) conversion directions between source and target speakers, many-to-many means multiple source speakers and multiple target speakers (3) number of parameters (4) autoregressive Emb denotes speaker embedding model.} 
    \label{tab:compare_vc}
\end{table*}

Table \ref{tab:compare_vc} presents an overview of cascade ASR-TTS \cite{Huang2020} and the proposed model. \cite{Huang2020} has more than 100 M (million) parameters in total. It is also notable that cascade ASR-TTS only supports many-to-one conversion direction, which means multiple source and only one target speaker are supported in each model. Hence, in scenarios with multiple target speakers, the parameter efficiency of cascade ASR-TTS is lower than models that support many-to-many conversion direction.

This paper focuses on improving efficiency of cascade ASR-TTS \cite{Huang2020}. Instead of using AR models, DYGAN-VC has a non-AR model structure, which is supposed to have better decoding efficiency. %It has been shown that non-AR VC models suffer from low voice quality issues \cite{Yi2020}. To improve voice quality, an additional discriminator is introduced. As shown in \cite{chou2018multi}, adversarial training with a discriminator enhances voice quality for VC.

Instead of using the Transformer ASR model, this paper proposes to use VQWav2vec.
VQWav2vec \cite{baevski2019vq} is one of speech self-supervised learning models \cite{schneider2019wav2vec,baevski2020wav2vec,hsu2021hubert} that encodes speech to features. VQWav2vec aims to learn unsupervised speech representations that benefit multiple downstreaming tasks. Based on Wav2vec \cite{schneider2019wav2vec}, a vector-quantization \cite{van2017neural} module is introduced, which is a differentiable clustering method. With the discreteness introduced to the model, VQWav2vec features are supposed to contain speech content information and also be speaker-invariant. %For example, in the recent SUPERB benchmark \cite{yang2021superb}, VQWav2vec representations achieved a 66.52 $\%$ accuracy in phone classification and a 38.8 $\%$ accuracy in speaker classification.
\begin{figure}[t]
    \centering
    \includegraphics[width=3in]{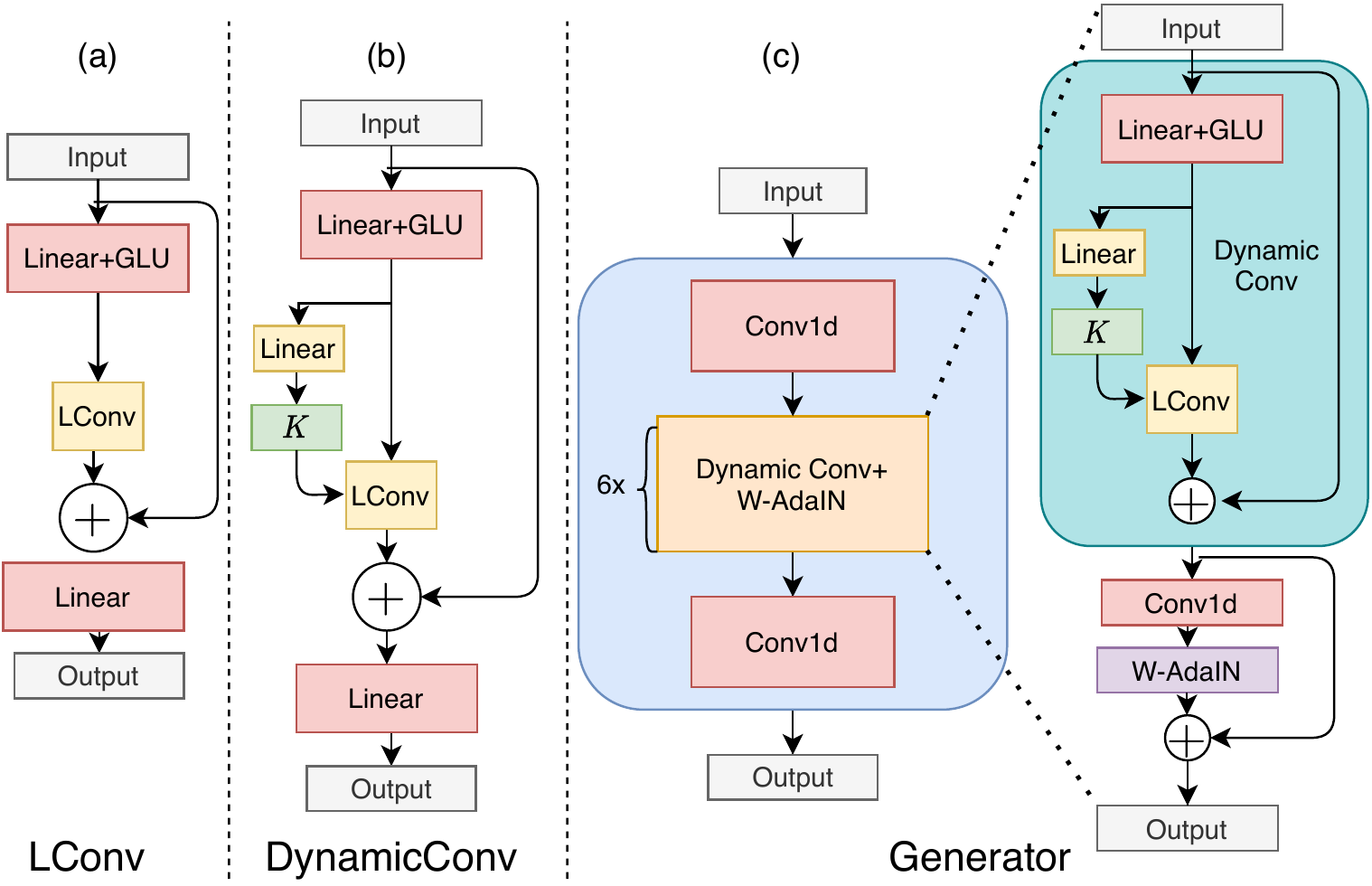}
    \caption{Architecture shown are (a): Lightweight convolution (LConv), (b) Dynamic convolution (DynamicConv), (c) is the Generator. K denotes dynamic convolution kernels}
    \label{fig:dyconv_gen}
\end{figure}
As shown in Table \ref{tab:compare_vc}, comparing to the ASR model used in cascade AST-TTS \cite{Huang2020}, as a non-AR model, VQWav2vec is smaller. Additionally, a recent VC work \cite{huang2021any} used VQWav2vec features to improve data efficiency.
 
To improve efficiency of the Transformer TTS model, instead of using computational costly self-attention layers, this paper proposes to use dynamic convolution \cite{wu2018pay} as a replacement.
Lightweight convolution and dynamic convolution \cite{wu2018pay} are proposed on the purpose of improving efficiency of large Transformer models. They can be seen as lightweight replacements for the computational expensive attention mechanisms of Transformer. Moreover, in a recent speech synthesis work \cite{elias2021parallel}, lightweight convolution has been introduced for better parameter and decoding efficiency.

This paper proposes DYGAN-VC, a novel VC model based on generative adversarial networks (GAN) \cite{goodfellow2014generative}. Instead of using large ASR models as in \cite{Huang2020}, DYGAN-VC uses VQWav2vec, which is lightweight. As a replacement of the self-attention layer, dynamic convolution \cite{wu2018pay} is introduced for better parameter efficiency. DYGAN-VC has a small model size and a fast decoding speed and reaches comparable performance to cascade ASR-TTS. The contributions can be summarized as follows
\begin{itemize}
    \item This paper proposes DYGAN-VC, an efficient GAN VC model with comparable performance of SOTA.
    \item This paper is the first to combines self-supervised features (VQWav2vec) with a GAN VC model.
    \item This paper is the first to introduce dynamic convolution to VC, which improves parameter efficiency.
\end{itemize}

\section{Background}
% introducing background
This section introduces background information for DYGAN-VC. It firstly introduces a comparison of lightweight convolution and dynamic convolution \cite{wu2018pay}. Then it introduces the differences of AdaIN \cite{huang2017arbitrary} and WadaIN \cite{kaneko2019stargan}, which is used in DYGAN-VC. 
%\subsection{VQWav2vec}

%VQWav2vec has been used in a recent VC model \cite{huang2021any} and been shown as good speech content representations. 
% \begin{table}[h]
%     \centering
%     \begin{tabular}{c|c|c|c}
%     Model     & \#Params & Phone ($\%$) & Speaker ($\%$) \\
%     \hline
%     Wav2vec     & 32,54M & 68.42 & 56.56 \\
%     VQWav2vec & 34.15M & 66.52 & 38.8\\
%     Wav2vec 2.0 & 95.04M & 94.26 & 75.18 \\
%     \end{tabular}
%     \caption{A comparison of model size, phone classification and speaker classification results of self-supervised models, results are from \cite{yang2021superb}.}
%     \label{tab:compare_ssl}
% \end{table}

%VQWav2vec encodes every 30 ms of waveforms to one vector with a stride size of 10 ms. It is trained by using a contrastive objective \cite{oord2018representation}, where the model tries to predict the next few steps features based on the previous features.

\subsection{Lightweight convolution and dynamic convolution}
% introduction of lightweight convolution ans dynamic convolution.
\label{sec:lconv_dyconv}
Lightweight convolution is a variant of 1d convolution, it has fewer parameters than vanilla 1d convolution. Given a feature matrix $X \in \mathbb{R}^{b \times t \times c}$, where $b, t, c$ denote the batch size, the segment length and the number of channels. Lightweight convolution has kernels $K \in \mathbb{R}^{k\times h}$, where $k$ is the kernel size and $h$ is the number of heads. The output $O \in \mathbb{R}^{b \times t \times c}$ is obtained by 
\begin{equation}
    o_{i,j,p} =  \sum_{q=1}^{k}  K_{q,(\ceil*{\frac{pc}{h}})} \cdot X_{i,(j+q - \ceil*{\frac{k+1}{2}}),p},
\end{equation}
where $o_{i,j,p}$ is an element of $O \in \mathbb{R}^{b \times t \times c} $.

Lightweight convolution splits the feature dimension of $X$ to $h$ groups, where features in one group share one kernel. By doing this, the number of parameters of one lightweight convolution layer is $k \times h$, which is less than a traditional 1d convolution layer. 

Based on lightweight convolution, dynamic convolution introduces an additional kernel generation mechanism that generates kernels from input features $X$, so that the shape of the kernels $K'$ for dynamic convolution becomes $[b,t,k,h]$. 

The following shows the formation of the kernel generation mechanism.
After a linear layer and a GLU layer, the feature matrix $X'$ can be obtained
\begin{equation}
    X' = GLU(XW_1+b_1).
\end{equation}

The dynamic convolution kernel $K' \in \mathbb{R}^{b \times t \times k \times h}$ can be generated through a linear layer.
\begin{equation}
    K' = X'W_2+b_2.
\end{equation}
The output of dynamic convolution can be obtained by using the generated kernel $K'$ and the feature $X$
\begin{equation}
    o'_{i,j,p} =  \sum_{q=1}^{k}  K'_{i,j,q,(\ceil*{\frac{pc}{h}})} \cdot X_{i,(j+q - \ceil*{\frac{k+1}{2}}),p},
\end{equation}
where $W_1 \in \mathbb{R}^{c \times (2*c)}$, $b_1 \in \mathbb{R}^{2*c}$, $W_2 \in \mathbb{R}^{c \times (k*h)}$, $b_2 \in \mathbb{R}^{k*h}$ are trained parameters. $o'_{i,j,p}$ is an element of the output $O' \in \mathbb{R}^{b \times t \times c}$.

The dynamic kernel generation mechanism produces a kernel for each time step ($K' \in \mathbb{R}^{b \times t \times k \times h}$), instead of using one kernel ($K \in \mathbb{R}^{k \times h}$) across all time steps. Hence, dynamic convolution gains a better ability for modeling local dynamic information, such as speech content.

\subsection{AdaIN and WadaIN}
\label{sec:adain_wadain}
% wadain layer.
\begin{figure}[t]
    \centering
    \includegraphics[width=3in]{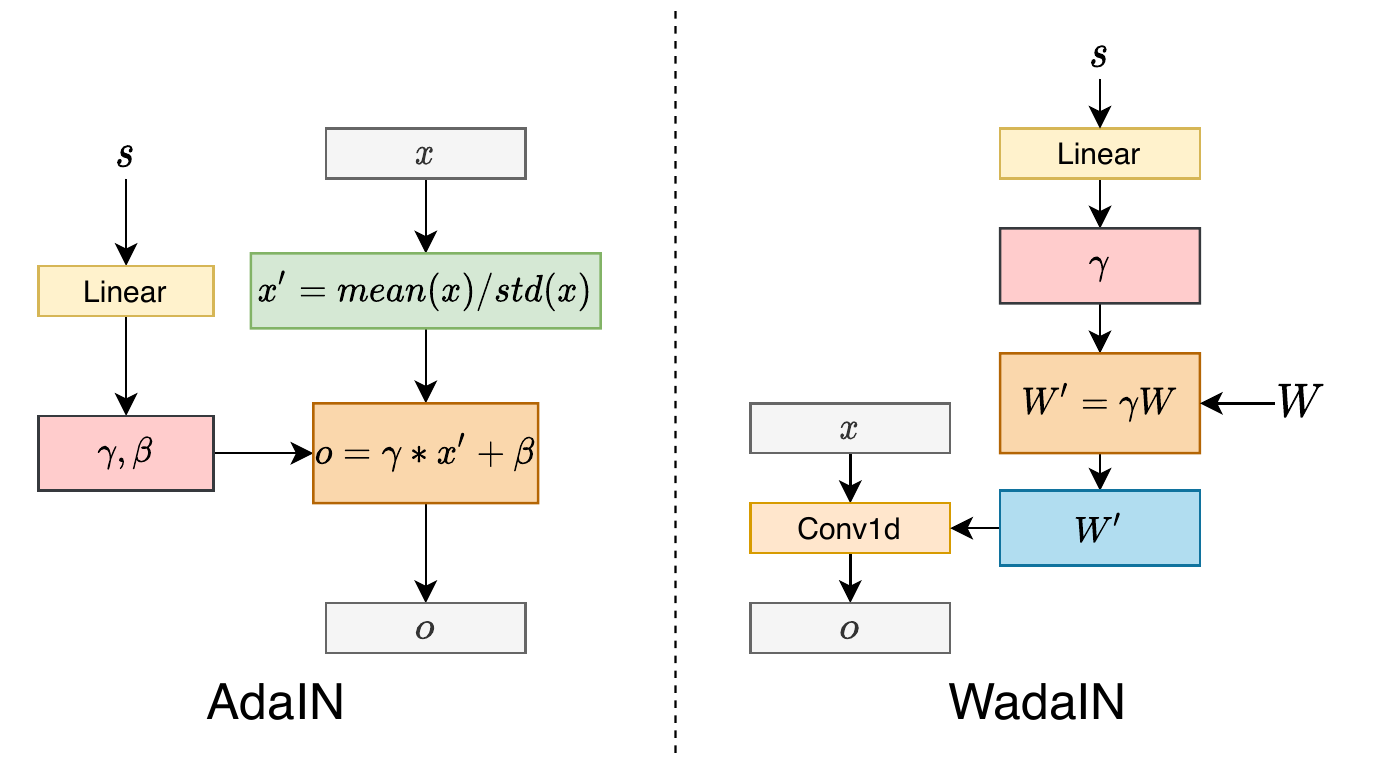}
    \caption{A comparison between AdaIN and WadaIN, where $x$ and $o$ denotes inputs and outputs, $\gamma$ and $\beta$ are affine parameters, $W$ denotes convolution kernels, $s$ is speaker embeddings }
    \label{fig:compare_adain_wadain}
\end{figure}
\begin{figure}[t]
    \centering
    \includegraphics[width=3in]{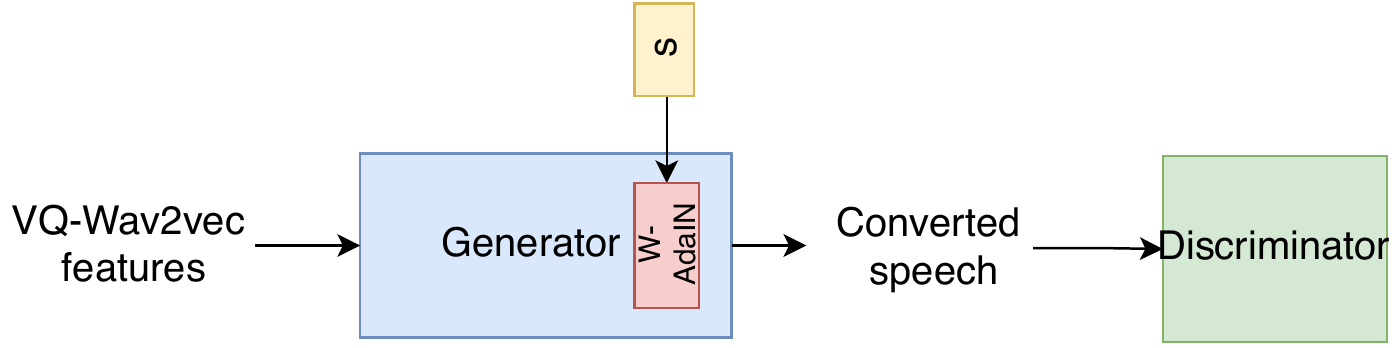}
    \caption{Overview of DYGAN-VC. The model is composed of a generator and a discriminator, s denotes target speaker embeddings.}
    \label{fig:overview}
\end{figure}
Figure \ref{fig:compare_adain_wadain} shows a comparison between AdaIN \cite{huang2017arbitrary} and WadaIN \cite{karras2020analyzing}. Given features $x$ and target speaker embeddings $s$, AdaIN adapts features $x$ to target speakers. It normalizes $x$ across the time dimension. The normalized features $x'$ are transformed with affine parameters $\beta$ and $\gamma$. The affine parameters are produced from target speaker embeddings $s$ through a linear layer.

\cite{karras2020analyzing} analysed the effects of AdaIN for a image generation task, where AdaIN was proved to cause artificials in the generated images. In a recent VC system \cite{chen2021towards}, WadaIN also showed advantages on voice quality.
As a replacement of AdaIN, WadaIN does not normalize or adapt features. It rather scales the channel dimension of convolution kernels. As shown in Figure \ref{fig:compare_adain_wadain}, convolution kernels $W$ are adapted to target speakers by a linear transform. The adapted kernels $W'$ can be obtained as follows:
\begin{equation}
    W' = \gamma * W,
\end{equation}
where $\gamma$ is an affine parameter generated from target speaker embeddings $s$ through a linear layer. The adapted convolution kernels $W'$ are used in a convolution layer where the input features are $x$.

\section{DYGAN-VC: a lightweight GAN model for voice conversion}

This section introduces the model architecture and the training objectives of DYGAN-VC. %More specifically, DYGAN-VC is a GAN \cite{goodfellow2014generative} model that takes in pretrained VQWav2vec features as inputs. DYGAN-VC proposes a novel generator architecture, which combines dynamic convolution \cite{wu2018pay} and WadaIN \cite{karras2020analyzing}.
As shown in Figure \ref{fig:overview}, DYGAN-VC is composed of a generator and a discriminator. The generator takes VQWav2vec as input and generates speech samples according to target speaker embeddings. The discriminator takes in the generated speech samples and returns a probability of the generated speech samples being real speech. The generator and the discriminator are trained adversarially. At inference time, only the generator is used.

\subsection{A new generator architecture: a combination of dynamic convolution and WadaIN}

Figure \ref{fig:dyconv_gen} (c) shows the architecture of the generator. It uses 1d convolution as the input layer and also as the ouput layer.
%The input convolution layer has a kernel size of 5 and the output convolution layer has a kernel size of 1. 
The generator contains 6 identical intermediate blocks. Each block includes a dynamic convolution layer, a 1d convolution layer and a WadaIN layer. In each intermediate block, layer normalization \cite{ba2016layer} and residual connections are applied as in \cite{vaswani2017attention}.

The novel intermediate block in the generator is inspired by AdaSpeech \cite{chen2020adaspeech}, which was proposed for a speaker adaptation task for TTS based on a Transformer TTS model \cite{ren2020fastspeech}.
%DYGAN-VC proposes a new fundamental block for the generator, which combines dynamic convolution \cite{wu2018pay} and WadaIN \cite{karras2020analyzing}. 
In AdaSpeech, each self-attention layer is followed by a variant of the AdaIN layer \cite{huang2017arbitrary}. Different from the original Transformer model \cite{vaswani2017attention} which applies linear layers after the self-attention layers, the design of AdaSpeech enables the Transformer model to adapt features to target speakers. More importantly, this design also maintains the voice quality of generated speech samples.
Hence, DYGAN-VC follows this idea and proposes to combine dynamic convolution layers \cite{wu2018pay} with WadaIN layers \cite{karras2020analyzing}.
Considering parameter efficiency, DYGAN-VC uses dynamic convolution instead of heavy self-attention layers. Moreover, as an extension of AdaIN, WadaIN enhances voice quality. By using this new combination, DYGAN-VC achieves both parameter efficiency and voice quality.
\subsection{Discriminator architecture}
The discriminator uses the same architecture as in StarGANv2-VC \cite{li2021starganv2}, except the maximum hidden size is reduced from 512 to 128. The discriminator has a input 2d convolution layer, which is followed by 4 residual convolution blocks. Each residual convolution block has two 2d convolution layers and an average pooling layer for downsampling. After 4 residual convolution blocks, there is a 2d convolution layer, a global average pooling layer and a 2d convolution layer. For more details, please refer to \cite{li2021starganv2}.

\subsection{Training objectives}
Given VQWav2vec features $z \in Z$ and speaker embeddings $s \in S$, the generator $G(z,s)$ generates converted speech samples $\hat{x} \in X$. The discriminator $D(x)$ takes in real speech samples $x \in X$ and the converted samples $\hat{x}$ and returns probabilities $y \in [0,1]$ of the input being real speech. 

The reconstruction loss can be defined as follows:
\begin{equation}
    \mathcal{L}_{recon} = \mathbb{E}_{x,z,s}||x - G(z,s)||.
\end{equation}
The least-square adversarial losses are used as in \cite{kaneko2019stargan}. The losses for the generator and the discriminator can be defined as follows:
\begin{equation}
   \mathcal{L}_{adv}^G = \mathbb{E}_{z,s}[( 1 - D(G(z,s)))^2],
\end{equation}
\begin{equation}
   \mathcal{L}_{adv}^D = \mathbb{E}_{x}[(1 - D(x))^2] + \mathbb{E}_{z,s}[ D(G(z,s))^2].
\end{equation}
The overall losses for the generator and the discriminator are as follows:
\begin{equation}
    \mathcal{L}^D = \mathcal{L}_{adv}^D ,
\end{equation}
\begin{equation}
    \mathcal{L}^G = \lambda \mathcal{L}_{recon} + \mathcal{L}_{adv}^G ,
\end{equation}
where $\lambda$ is a hyper-parameter set as 5.
% \begin{table*}[t]
    
%     \centering
%     \begin{tabular}{c|c|c|c|c}
%     Model     & F-F & F-M & M-F& M-M\\
%     \hline
%      {} & MCD/MOS/CER/WER & MCD/MOS/CER/WER & MCD/MOS/CER/WER&MCD/MOS/CER/WER  \\
%      \hline
%      ASR-TTS & 6.65/ 3.87/5.4/10.1& 6.19/3.42/9.1/15.0&6.62/3.86/5.8/10.7&6.14/3.42/8.1/13.6\\
%      StarGANv2-VC &8.34/ 4.28/ 5.2/10.7 & 7.61/3.60/4.2/ 8.9&8.29/4.34/3.7/7.8&7.55/3.63/4.0/8.3 \\
%      %CycleGAN-VC3& 8.70/3.77/4.7/10.27& 8.60/3.53/26.2/47.0&9.16/3.68/15.7/28.6&7.90/3.58/6.3/12.6\\
%      %CycleVQVAE& 8.02/4.30/15.2/27.3 & 8.73/3.60/25.6/43.5&8.15/3.49/18.3/33.6&7.26/3.49/14.8/25.1  \\
%      DYGAN-VC & 7.56/4.33/6.1/12.6&7.01/3.85/5.9/11.9&7.54/4.26/4.8/10.9&7.02/3.79/4.0/9.3\\
%      \hline
%     \end{tabular}
    
%     \caption{Objective evaluation results. F-F, F-M, M-F, M-M denote conversion directions, F means female, M means male. MCD, MOS, CER, WER denote mel-cepstral distorsion, MOSNet, character error rate ($\%$) and word error rate ($\%$), respectively. Models presented in this table have huge differences, please refer to Table \ref{tab:compare_vc} for details.}
%     \label{tab:obj_res}
% \end{table*}
\begin{table*}[t]
    
    \centering
    \begin{tabular}{c|c|c|c|c}
    Model  & F-F & F-M & M-F& M-M\\
    \hline
     {} &MCD/MOS/CER/WER & MCD/MOS/CER/WER & MCD/MOS/CER/WER&MCD/MOS/CER/WER  \\
     \hline
     Cascade ASR-TTS & 6.65/ 3.87/5.4/10.1& 6.19/3.42/9.1/15.0&6.62/3.86/5.8/10.7&6.14/3.42/8.1/13.6\\
     %StarGANv2-VC &8.34/ 4.28/ 5.2/10.7 & 7.61/3.60/4.2/ 8.9&8.29/4.34/3.7/7.8&7.55/3.63/4.0/8.3 \\
     %CycleGAN-VC3& 8.70/3.77/4.7/10.27& 8.60/3.53/26.2/47.0&9.16/3.68/15.7/28.6&7.90/3.58/6.3/12.6\\
     %CycleVQVAE& 8.02/4.30/15.2/27.3 & 8.73/3.60/25.6/43.5&8.15/3.49/18.3/33.6&7.26/3.49/14.8/25.1  \\
     DYGAN-VC & 7.56/4.25/5.5/11.2&7.04/3.77/6.2/13.4&7.52/4.20/4.3/10.2&7.03/3.78/4.5/10.7\\
     \hline
    \end{tabular}
    
    \caption{Objective evaluation results. F-F, F-M, M-F, M-M denote conversion directions, F means female, M means male. MCD, MOS, CER, WER denote mel-cepstral distorsion, MOSNet, character error rate ($\%$) and word error rate ($\%$), respectively. Please note that models presented in this table have large differences, please refer to Table \ref{tab:compare_vc} for details.}
    \label{tab:obj_res}
\end{table*}
\begin{table*}[t]
    \centering
    \begin{tabular}{c|c|c|c|c}
    Model & F-F& F-M& M-F& M-M  \\
    \hline
          & Nat/Sim&Nat/Sim&Nat/Sim&Nat/Sim\\
    \hline
    Cascade ASR-TTS & 3.81$\pm$0.12/3.70$\pm$0.12&3.81$\pm$0.12/3.84$\pm$0.12&3.90$\pm$0.12/3.83$\pm$0.14&3.82$\pm$0.12/3.83$\pm$0.16 \\
    %StarGANv2-VC& 3.84$\pm$0.12/3.87$\pm$0.13&3.83$\pm$0.11/3.76$\pm$0.12&3.96$\pm$0.13/4.01$\pm$0.14&3.90$\pm$0.14/3.91$\pm$0.14  \\
    DYGAN-VC & 3.80$\pm$0.13/3.79$\pm$0.13&3.74$\pm$0.13/3.84$\pm$0.13&3.86$\pm$0.13/3.94$\pm$0.13&3.83$\pm$0,13/3.92$\pm$0.12 \\
    \hline
         
    \end{tabular}
    \caption{Subjective evaluation results with $95\%$ confidential intervals, Nat and Sim denote naturalness and speaker similarity. Please note that models presented in this table have large differences, please refer to Table \ref{tab:compare_vc} for details.}
    \label{tab:subject_res}
\end{table*}
\begin{table}[t]
    \centering
    \begin{tabular}{c|c|c|c|c}
         &MCD & MOSNet& CER$\%$& WER$\%$  \\
         \hline
    lconv+AdaIN & 7.36&3.89&5.4&11.5 \\ 
    dyconv+AdaIN& 7.35& 3.96& 6.3&12.8 \\
    lconv+WadaIN&7.29&\textbf{4.01}&6.1&12.4 \\
    dyconv+WadaIN & \textbf{7.29}&4.00&\textbf{5.2}&\textbf{11.2}\\
    \hline
    \end{tabular}
    \caption{A comparison between options of dynamic convolution, lightweight convolution, AdaIN and WadaIN, results are averaged over all 16 conversion pairs.}
    \label{tab:ablation}
\end{table}
\section{Experimental setup}
\label{sec:experiment_setup}
\subsection{Dataset}
This work uses the first track dataset of VCC2020 \cite{Yi2020}. The dataset contains 8 English speakers, including 4 female speakers and 4 male speakers. In average, each speaker has 5 minutes of data, so the whole dataset contains 40 minutes of audio data. The speakers are composed of 4 sources speakers (2 female, 2 male) and 4 target speakers (2 female, 2 male), so there are $4 \times 4 = 16$ conversion pairs in total. There are 70 training samples and 25 testing samples for each speaker. This work uses 60 samples for training, 10 samples for validation and 25 samples for testing.
\subsection{Baseline models}
The Cascade ASR-TTS model \cite{Huang2020} is the official baseline model of VCC2020 \cite{Yi2020}. This work uses the official recipe \footnote{\url{https://github.com/espnet/espnet/tree/master/egs/vcc20/vc1_task1}} as implementation. %Additionally, 
%StarGANv2-VC \cite{li2021starganv2} is a recent SOTA GAN VC model. This work uses the official implementation \footnote{\url{https://github.com/yl4579/StarGANv2-VC}}.% CycleGAN-VC3 \cite{kaneko2020CycleGAN-VC3} is a non-AR model and does not depend on pretraining resources. This work chooses CycleGAN-VC3 as its model size (27 M) as DYGAN-VC (42 M). 
\subsection{Evaluation}
Objective evaluations and subjective evaluations are conducted. Four objective metrics are used: mel-cepstrum distortion (MCD), character error rate (CER), word error rate (WER) and MOSNet \cite{lo2019mosnet}. The MCD, CER and WER results were calculated using the official baseline implementation \cite{Huang2020}. The MOSNet score was calculated using crank \cite{kobayashi2021crank}.

Mean opinion score (MOS) human evaluations were conducted as subjective evaluations. The generated speech samples were evaluated for naturalness and similarity. Listening tests were conducted on the MTurk \footnote{\url{https://www.mturk.com}} platform. For naturalness, listeners were asked to mark a speech sample at five (1-5) grades. For speaker similarity, the listeners were asked to mark a speech sample at five (1-5) grades  given a reference speech sample. In order to recognize listeners with bad behaviours, ground truth speech samples and random noise samples were included in test sets.
After removing 2 outlier listeners, 192 listeners participated the listening tests.
% todo
The test set contains 400 samples for each model, resulting in 1200 samples in total. Each sample is evaluated by at least 2 listeners for both naturalness and speaker similarity.
\subsection{Implementations}
80 dimensional mel-spectrograms are used as features, window size is 40 ms and hop size is 10 ms. The vocoder model is the Parallel WaveGAN model \cite{yamamoto2020parallel}. The learning rate for the generator and the discriminator is 1e-4 and 2-e5, respectively. Batch size is 8, and speaker embeddings are extracted using the speaker encoder model in \cite{liu2021any}. At training time, speech utterances are cropped to segments of 128 frames. Adam \cite{kingma2014adam} is used as optimizers. The model is trained for 100 epochs and the training takes about 40 minutes to converge on one gpu. For more details, please refer to implementation \footnote{Source code implementation can be found in \url{https://github.com/MingjieChen/DYGANVC}, demo page \url{https://mingjiechen.github.io/dygan-vc/}}.

% \begin{table*}[t]
%     \centering
%     \begin{tabular}{c|c|c|c|c|c}
%     Model &\#Param& F-F& F-M& M-F& M-M  \\
%     \hline
%          & & Nat/Sim&Nat/Sim&Nat/Sim&Nat/Sim\\
%     \hline
%     ASR-TTS&103M & 3.81$\pm$0.12/3.70$\pm$0.12&3.81$\pm$0.12/3.84$\pm$0.12&3.90$\pm$0.12/3.83$\pm$0.14&3.82$\pm$0.12/3.83$\pm$0.16 \\
%     StarGANv2-VC&82M& 3.84$\pm$0.12/3.87$\pm$0.13&3.83$\pm$0.11/3.76$\pm$0.12&3.96$\pm$0.13/4.01$\pm$0.14&3.90$\pm$0.14/3.91$\pm$0.14  \\
%     DYGAN-VC &42M& 3.80$\pm$0.13/3.79$\pm$0.13&3.74$\pm$0.13/3.84$\pm$0.13&3.86$\pm$0.13/3.94$\pm$0.13&3.83$\pm$0,13/3.92$\pm$0.12 \\
%     \hline
         
%     \end{tabular}
%     \caption{Subjective evaluation results with $95\%$ confidential intervals, N and S denote naturalness and speaker similarity.}
%     \label{tab:my_label}
% \end{table*}
% \begin{table}[t]
%     \centering
%     \begin{tabular}{c||c|c|c|c}
%          &MCD & MOS& CER$\%$& WER$\%$  \\
%          \hline
%     lconv+Adain & 7.36&3.89&5.4&11.5 \\ 
%     dyconv+Adain& 7.35& 3.96& 6.3&12.8 \\
%     lconv+WadaIN&7.29&4.01&6.1&12.4 \\
%     dyconv+WadaIN & \textbf{7.28}&\textbf{4.06}&\textbf{5.2}&\textbf{11.1}\\
%     \hline
%     \end{tabular}
%     \caption{A comparison between options of dynamic convolution, lightweight convolution, AdaIN and WadaIN, results are averaged over all 16 conversion directions}
%     \label{tab:ablation}
% \end{table}
\section{Results}

\subsection{Objective results}

Table \ref{tab:obj_res} demonstrates objective evaluation results for four conversion directions. Generally speaking, considering all four directions, the performance of DYGAN-VC is at the same level of the baseline model. %Cacade ASR-TTS leads the MCD scores. As for MOSNet, DYGAN-VC achieves leading scores. StarGANv2-VC has the best results for CER and WER.   
%Note that Cacade ASR-TTS model has advantages on model size and it is an AR model. StarGANv2-VC is a non-AR model but still has advantages on model size.
Comparing DYGAN-VC with cascade ASR-TTS, the latter model has better MCD results for all directions. This might because cascade ASR-TTS is composed of AR models, hence it has better ability to model target speaker properties. DYGAN-VC has better MOSNet scores for all directions. DYGAN-VC has better CER and WER results for F-M, , M-F and M-M directions but worse results for F-F directions. 
%Comparing DYGAN-VC with StarGANv2-VC, StarGANv2-VC generally has better CER and WER results. This might because of benefits of the huge model size of StarGANv2-VC. DYGAN-VC has much better MCD results for all directions. Also, DYGAN-VC has better results of MOSNet for F-M and M-M directions but worse for F-F and F-M directions.
%Overall, considering the differences of model sizes and whether the model is an AR model, DYGAN-VC achieves the same level of objective results with better model efficiency.  
\subsection{Subjective results}
Table \ref{tab:subject_res} demonstrates subjective evaluation results for all conversion directions. %StarGANv2-VC leads naturalness MOS scores for all directions. Also StarGANv2-VC leads speaker similarity MOS scores except F-M direction. 
Comparing DYGAN-VC with cascade ASR-TTS, for naturalness MOS scores, cascade ASR-TTS has better results than DYGAN-VC except M-M direction. However, the distance of naturalness MOS scores between them is within 0.1, for example, for F-F direction, DYGAN-VC achieves 3.80, which is very close to 3.81 of cascade ASR-TTS. As for speaker similarity, DYGAN-VC has better results than Cascade ASR-TTS except F-M direction.% Similar to naturalness, the distance of results between two models in terms of speaker similarity is small.
%Comparing DYGAN-VC with StarGANv2-VC, for naturalness MOS scores, StarGANv2-VC has better results for all directions. For speaker similarity MOS scores, StarGANv2-VC has better results for F-F and M-F directions but worse results for F-M and M-M directions. 
\subsection{Ablation study}
To further study the effects of the proposed new combination of dynamic convolution and WadaIN, this work compares objective results for all possible combinations, as shown in Table \ref{tab:ablation}.
The combination of dynamic convolution and WadaIN reaches the best results for MCD, CER and WER, except MOSNet.  
\subsection{Decoding speed}
\begin{table}[t]
    \centering
    \begin{tabular}{c|c}
    Model & RTF \\
    \hline
    Cascade ASR-TTS & 46.1\\ 
    %StarGANv2-VC & 10.4\\
    DYGAN-VC & 5.4 \\
    \hline
    \end{tabular}
    \caption{A comparison of the decoding speed, RTF denotes real time factor, which is the average time for generating one second of waveform on cpu.}
    \label{tab:speed}
\end{table}

Table \ref{tab:speed} compares the decoding speed on cpu at inference time. This work reports the real time factor (RTF) as decoding speed. Since cascade ASR-TTS is an AR model, it has a slower decoding speed than non-AR DYGAN-VC.
\section{Conclusion}
This paper proposes DYGAN-VC, a novel GAN VC model with high efficiency. Instead of using a ASR model, DYGAN-VC uses VQWav2vec, which is lightweight and faster. Furthermore, DYGAN-VC introduces dynamic convolution, which enhances speech content modeling and keeps a lightweight model. Comparing to the SOTA models, DYGAN-VC has high efficiency, and also achieves comparable level of performance of SOTA. For future work, the authors will investigate zero shot VC.

\newpage
\bibliographystyle{IEEEtran}

\bibliography{mybib}

\end{document}